\documentclass[runningheads]{llncs}
\usepackage{graphicx}
\usepackage{multirow}
\usepackage{booktabs}
\usepackage[misc]{ifsym}
\usepackage{array}
\usepackage{amssymb}
\usepackage{verbatim}
\usepackage{array}

\begin{document}
\title{Pulmonary Vessel Segmentation based on Orthogonal Fused U-Net++ of Chest CT Images}

\author{Hejie Cui \inst{1, 2}  \and
Xinglong Liu\inst{1 } \and
Ning Huang\inst{1 ( \textrm{\Letter} )}}

%
\authorrunning{H. Cui et al.}
%
\institute{SenseTime Research, Beijing, China \\
\email{\{huangning\}@sensetime.com}
\and 
Emory University, Atlanta, GA, USA  }
\titlerunning{ }
%
%
%
%
\maketitle              
%

\begin{abstract}
Pulmonary vessel segmentation is important for clinical diagnosis of pulmonary diseases, while is also challenging due to the complicated structure. In this work, we present an effective framework and refinement process of pulmonary vessel segmentation from chest computed tomographic (CT) images. The key to our approach is a 2.5D segmentation network applied from three orthogonal axes, which presents a robust and fully automated pulmonary vessel segmentation result with lower network complexity and memory usage compared to 3D networks. The slice radius is introduced to convolve the adjacent information of the center slice and the multi-planar fusion optimizes the presentation of intra and inter slice features. Besides, the tree-like structure of pulmonary vessel is extracted in the post-processing process, which is used for segmentation refining and pruning. In the evaluation experiments, three fusion methods are tested and the most promising one is compared with the state-of-the-art 2D and 3D structures on 300 cases of lung images randomly selected from LIDC dataset. Our method outperforms other network structures by a large margin and achieves by far the highest average DICE score of 0.9272 and a precision of 0.9310, as per our knowledge from the pulmonary vessel segmentation models available in literature. 

\keywords{Pulmonary Vessel Segmentation  \and U-Net++ \and 2.5D CNN.}
\end{abstract}

\section{Introduction}

Pulmonary vessel segmentation is a topic of high interest in the field of medical image analysis: accurate vascular analysis has extremely important research and application value for treatment planning and clinical effect evaluation. Pulmonary vessel segmentation is a basis for common pulmonary vascular diseases diagnosis such as lobectomy and pulmonary embolism \cite{el2011lung}. 

However, the lung of the human body is the exchange place for metabolically produced gases, which is rich in trachea and vascular tissues, so its structure is relatively complicated. At the same time, due to factors such as noise and volume effect, CT images might suffer from poor contrast and blurred boundaries. Moreover, the pulmonary venous arteries and veins are intertwined and accompanied, which further increases the difficulty of segmentation \cite{lesage2009review}. 

A number of earlier vessel segmentation methods like tracking algorithms \cite{shikata2004automated}, seed point based \cite{kaftan2008fuzzy}, edge-based or region-based deformable model  \cite{staal2004ridge} have been applied and tested in different anatomical regions or imaging modalities, such as retina images. These methods, however, depend on hand-crafted features, thus having limited feature representation abilities. Besides, few supervised methods have been applied on pulmonary vessel segmentation. This is due to the inaccessibility of complex, fully-annotated dataset, which has become an important factor limiting the development of deep learning algorithms in this task \cite{rudyanto2014comparing}. Besides, relevant vessel segmentation studies have proposed the use of synthetic data for 2D or 3D neural network training \cite{schneider2012tissue}, but considering the complexity of real blood vessel distribution and the influence of pathological tissue variability, hardly can synthetic data truly and comprehensively reflect the pattern of vascular tree. 

Fully connected neural networks (FCNs) have achieved general success on segmentation tasks. In order to find an effective segmentation method for pulmonary vessel segmentation task, we do early-stage experiments on both 2D FCNs and 3D FCNs architecture with volumetric input. Result shows that 2D FCNs ignore the context information along the stacked axis, which contains important connection information for the upper and lower levels of the vascular tree; while 3D FCNs suffer from high computational cost and GPU memory consumption, which impedes the performance for large scale dataset  \cite{yun2019improvement}.

To better solve the problems mentioned above, we have proposed a fused 2.5D U-Net++ applied from three orthogonal axes. We use volumetric ground truth generated by unsupervised methods and then manually corrected by professional radiologists as the input of 2D convolutional network for each direction. The voxel prediction results of adjacent slices are employed for the prediction of the center slice, and segmentation volume of each axis is gained by stacking the segmentation maps of the center slices. This 2.5D convolution process is applied from three axes, where 3D contexts are effectively extracted and jointly optimized for an accurate pulmonary vessel tree segmentation. The evaluation of our proposed network achieves a reliably better result compared with several state-of-the-art segmentation networks. Besides, we also propose a post-processing process where the tree-like skeleton is generated for the refining of the segmentation result. Our contributions mainly lie in:

\begin{itemize}
\item We propose a 2.5D convolution network, which employs the 2D convolutional network on a stack of adjacent slices and fuses the features extracted from three orthogonal axes. 

\item A whole automated segmentation framework is given and we introduce a post-processing where the segmentation result is refined by the graph information of pulmonary vessel tree.

\item Our method gives a very competitive performance and ranks 1st till now on DICE Similarity Coefficient and Precision compared with the results reported by other state-of-the-art methods.
\end{itemize}

\section{Methods}
\begin{figure}
\includegraphics[width=\textwidth]{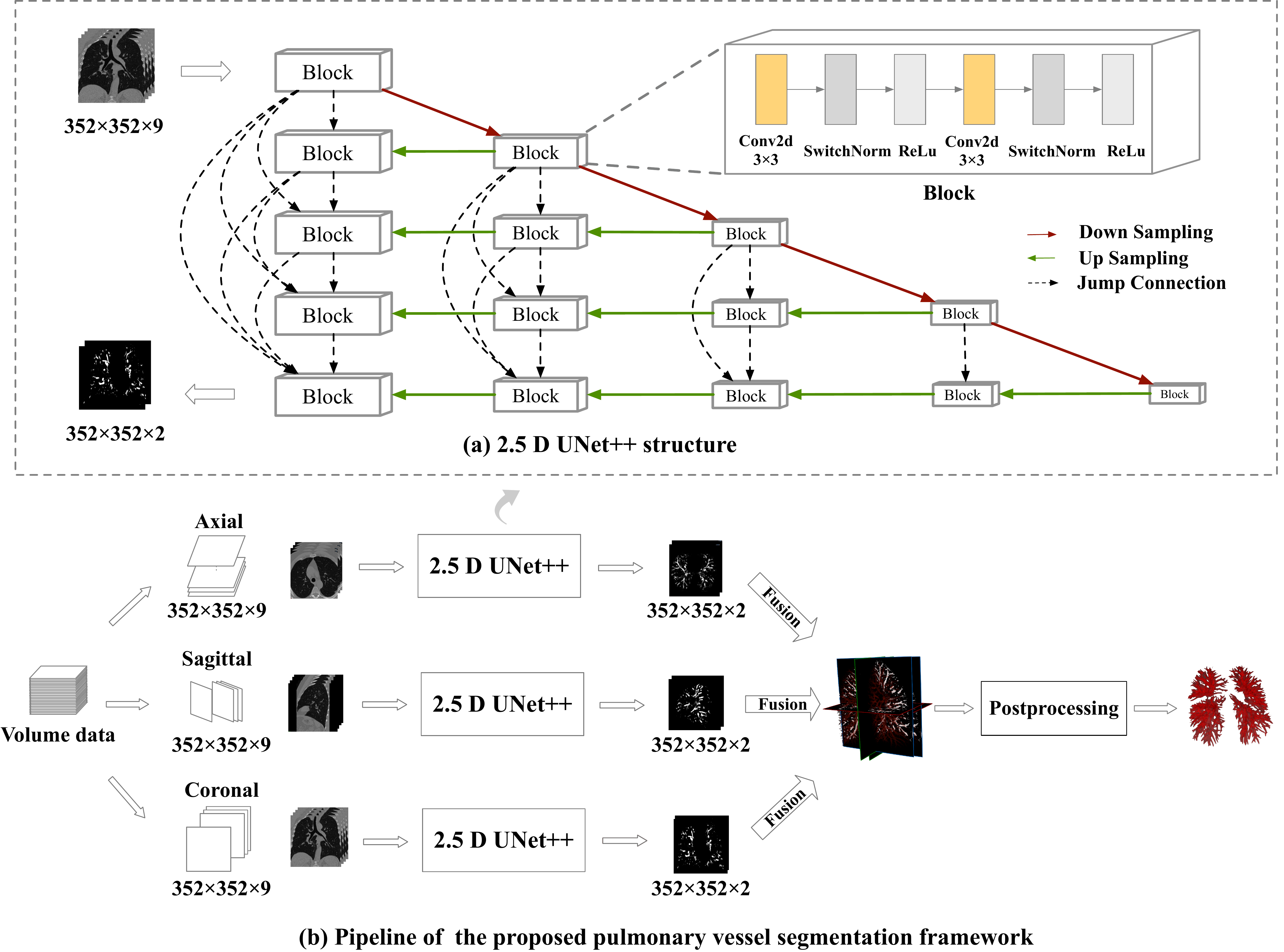}
\caption{Overall architecture of the proposed pulmonary vessel segmentation framework. Each volume data is sampled as adjacent slice groups along three orthogonal axes and then fed to a 2.5D U-Net++ network; The features extracted by the three parallel 2.5D networks are fused to optimize the volumetric representation; In the post processing, a structure graph is extracted to refine the segmentation result. (a) The structure of 2.5D U-Net++, whose output is the two-channel probability map of the center slice. (b) The whole framework, including the orthogonal fusion of multi-planar networks and the post-processing for segmentation refinement. } \label{fig1}
\end{figure}

We combine the idea of 2.5D network \cite{yun2019improvement} and orthogonal fusion of multi-planar network to obtain a new architecture: an orthogonal fused 2.5D U-Net++. Fig.~\ref{fig1} shows the pipeline of our proposed method for pulmonary vessel segmentation. 

\noindent\textbf{2.5D Network Based on U-Net++}
In the clinical practice, the experienced radiologists usually observe and judge the full structure of pulmonary vessel based on several successive slices along a specific axis. Therefore, a conventional 2D based network will easily ignore the context information while extracting the intra slice features, thus restricting the segmentation accuracy \cite{li2018h}.

In order to include the adjacency information between slices while doing the vessel segmentation, we design a 2.5D network which digests multiple channels cropped from the original CT images. Slice radius is introduced, where slices within the radius will be convolved during the feature map extraction process. The input channel is 9, and the ground-truth of the middle slice will be provided. The upper and lower 4 pieces of the middle slice are used to generate the feature maps. Since the information within radius range will be convolved, the inter slice information is preserved as much as possible to help the segmentation of middle slices. The output channel of the 2.5D network is 2, indicating the voxel-wise probability of being foreground or background. \\

\noindent\textbf{Orthogonal Fusion of Multi-planar Networks} 
Adjacent slices along each of the orthogonal directions including axial, sagittal, and coronal provide different connectivity information. To segment the candidate vessel voxels, the slice groups along three axes are processed in parallel. Slices within the radius will go through one of three separated identical up-sampling, down-sampling and convolution process \cite{zhou2018unet++}, which is composed by a stack of VGG blocks \cite{simonyan2014very}. These three parallel results of each direction are then fused under the comparison of different methods, including intersection, union and average value. By jointing together the intra and inter slice features extracted along three orthogonal axis, we optimize the description of volumetric feature representation to be more integral and comprehensive. The average fusion outperforms others on DICE, precision and recall value so we adopt it in the proposed orthogonal fusion model.\\

\noindent\textbf{Two-Stage Loss Function} 
For vessel segmentation task, the object of interest accounts far less than the background voxels in most cases, which leads to a high rate of false positive and recall values \cite{tetteh2018deepvesselnet}. To alleviate the class-imbalance problem caused by the inequitable penalty of positive and negative voxels, we separate the training process as two stage: 1) In the pre-trained stage, we use NLL(Negative Log Likelihood) loss to get a coarse model; 2) In the fine-tuned stage, we resume the coarse model and adopt a weight-balanced loss to suppress the over-segmentation and high false-positive rate: the calculated voxel-wised losses of both positive and negative positions are sorted, and the negative sorted list is much longer than the positive one considering the small occupancy of interest regions. We cut the negative list to make it the same length as the positive list. The top part of negative list is taken for loss function in order to balance the weight between the proportion of two kinds of voxels. The weighted loss function employed is described as below: \\


\vspace{-0.5cm}

\begin{eqnarray}
L &= &L _{Y_{+}(W)}+L _{Y_{-}(W)} \\
L _{Y_{+}(W)} &=& - \sum_{i\in{Y_{+}}}^{N}logP(y_{i}=1|X; W)\\
L _{Y_{-}(W)}& =& - \sum_{j\in{\widetilde{Y}_{-}}}^{N}logP(y_{j}=0|X;W)
\end{eqnarray}
where $\widetilde{Y}_{-}$ represents the top $N$ loss value selected from the sorted list of negative samples and $N$ is the number of elements in positive list.\\

\noindent\textbf{Vessel Structure Generation} 
In the post-processing stage, vessel structure is used to refine the segmentation result. We generate the morphology representation of tree-like graph from the skeleton of segmentation result. The graph includes nodes and edges, and the connected components can be calculated. This tree-like graph of pulmonary vessel can express plenty of potential useful information at very fine scales. In the post-processing, connected components with less than 10 nodes are trimmed on the graph, then the refined graph is filled edge by edge to get a refined segmentation result. The input of the post-processing is segmentation result of the end-to-end segmentation network and the output is refined vessel tree segmentation composed by the main connected components with more than 10 nodes \cite{shang2011vascular}. 

In addition to pruning the segmentation result, the topological structure represented by edges and nodes also indicates meaningful information for clinical practice, such as the location of junction points, the number of individual branches, and the connection relationship between bifurcations and end-points. We will continue to explore its application in future work.\\


\section{Experiments}
\noindent\textbf{Dataset and Pre-processing}
We randomly select a subset of 300 cases of chest CTs from publicly available LIDC  \cite{armato2011lung} dataset and split them into 270 cases for training and 30 cases for validating. 10\% of the selected dataset is utilized as testing set. To ensure the data variability, both challenging and visible vessels are included to cover a comprehensive situation. The ground truth mask for training is first generated using unsupervised method and then refined and validated by expert radiologists. 

Pre-processing includes two parts: resolution regularization and Hounsfield Unit (HU) Value normalization. The original resolution varies from $0.6\times0.6\times1.25 \textrm{ mm}^3$ to $0.9\times0.9\times2.5 \textrm{ mm}^3$. For resolution regularization, we resample the data to $1 \textrm{ mm}^3$ resolution cube. For intensity normalization, we adopt a lung window of $[-1200, 600]$ HU. The HU value of all data is cropped and adjusted to the range of lung window and then normalized to $[0, 1]$. \\

\noindent\textbf{Implementation Details}
The training of our model is performed on a workstation with a CPU of Intel(R) Core(R) i7-7700 @ 3.6 GHz and a NVIDIA GTX 1080 Ti GPU with 11GB of memory. In the training process of the fused 2.5D network, we use the SGD optimizer with a momentum of 0.99 and a weight decay of 1e-8. The initial learning rate is 0.001 and we apply a stepped learning rate scheduler with the initial value multiplied by a specific gamma value every several epochs. The loss function is divided into two stages: in the pre-trained stage, we use NLL-loss; and in the fine-tuned stage, we propose the weight-balanced loss to alleviate the disproportionate rate between the foreground and background. \\

\begin{table}[!t]
\caption{Comparison of Six Fusion Methods.}\label{tab1}
\centering
{\renewcommand{\arraystretch}{1.2}
\begin{tabular}{p{0.3cm}p{2cm}<{\centering}p{2cm}<{\centering}p{2cm}<{\centering}p{2cm}<{\centering}p{3.2cm}<{\centering}p{1cm}}
\toprule
\toprule
&Methods&Min Dice&Max Dice&Avg. Dice&Precision/Recall\\
\midrule
&Axial&0.8618&0.9584&0.9162&0.9250/0.9144\\
&Sagittal&0.8444&0.9575&0.9114&0.9232/0.9080\\
&Coronal&0.7474&0.9547&0.8964&0.9024/0.9020\\
&Union&0.8252&0.9580&0.9118&0.8714/0.9634\\
&Intersection&0.7946&0.9555&0.9096&0.9518/0.8772\\
&Average&\textbf{0.8779}&\textbf{0.9627}&\textbf{0.9262}&\textbf{0.9310/0.9272}\\
\bottomrule
\bottomrule
\end{tabular}
}
\end{table}

\noindent\textbf{Results and Discussion}
The evaluation experiments include two parts. 

First, we compare the single axis model with three fusion methods, including intersection, union and average value, to find the best one in keeping information of three axes. The pulmonary vessel segmentation result of each fusion method is presented in Table. \ref{tab1}. Grid search in the range of $(0.05, 0.5)$ is used to find the best threshold value of the prediction result map for each fusion method. We record the minimal, maximum, average dice value and precision/recall ratio under the best threshold of each method. Results show that the average fusion method achieves the highest performance in all statistical indicators. 

\vspace{-0.3cm}
\begin{figure}
\includegraphics[width=\textwidth]{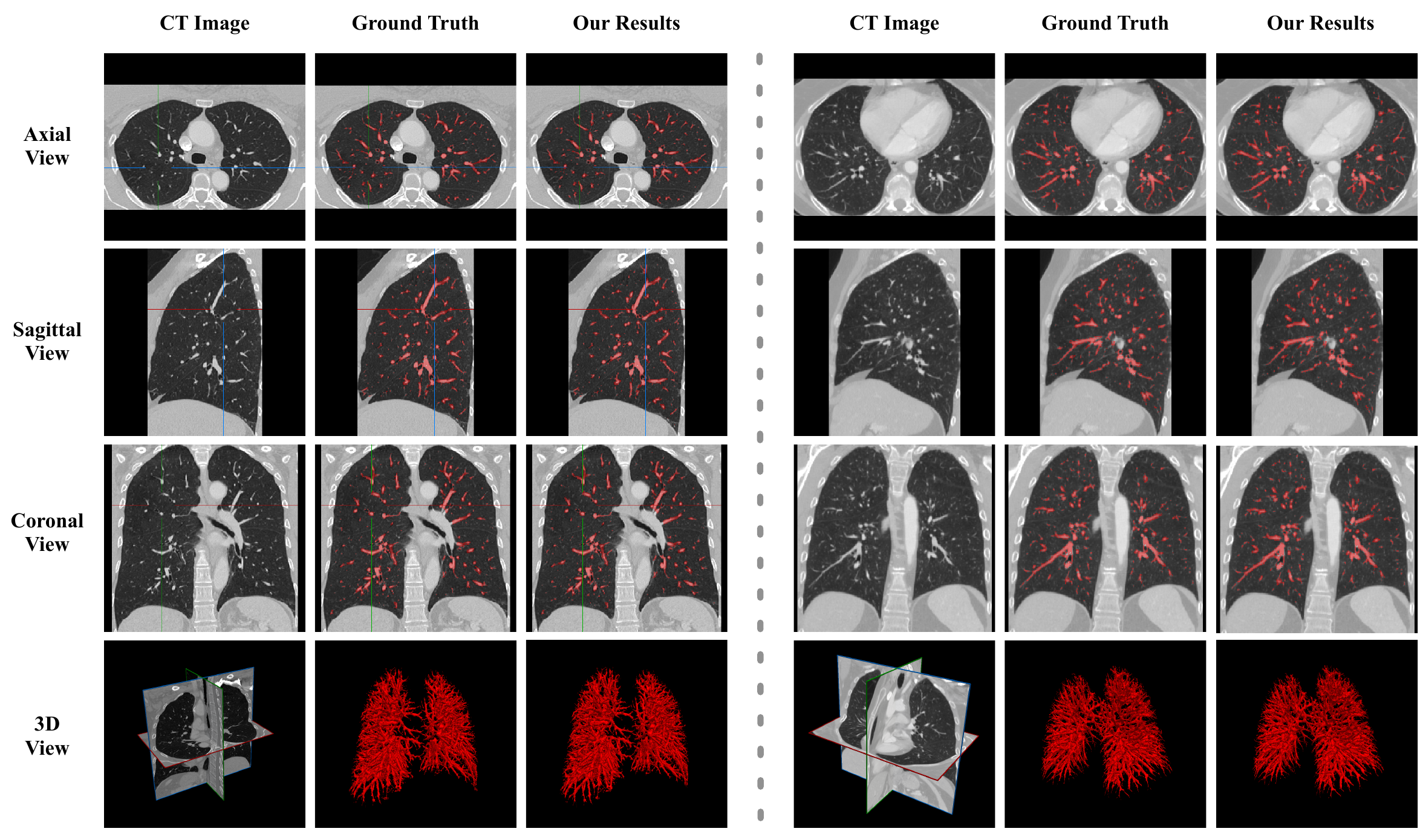}
\vspace{-0.4cm}
\caption{The visual comparison between the ground truth and result of the proposed 2.5D Average Orthogonal Fused U-Net++.} \label{fig2}
\end{figure}

Sencond, we further compare the average fused 2.5D network structure with several state-of-the-art network structures, including 2D U-Net++, 3D U-Net++ as well as several 3D FCNs. The quantitive results of U-Net++ based model are shown in Table. \ref{tab2} for comparison on a single-model basis. The promising 2.5D network significantly outperforms 2D and 3D FCNs models, which validates the advantage of our proposed structure. 

\begin{table}[!h]
\caption{Comparison of Three State-of-art Structures.}\label{tab2}
\centering
{\renewcommand{\arraystretch}{1.2}
\begin{tabular}{p{0.3cm}p{3cm}p{1.8cm}<{\centering}p{1.8cm}<{\centering}p{1.8cm}<{\centering}p{2.8cm}<{\centering}}
\toprule
\toprule
&Structures&Min Dice&Max Dice&Avg. Dice&Precision/Recall\\
\midrule
&2D U-net++&0.5201&0.7376&0.6628&0.6629/0.6767\\
&3D U-Net++&0.4385&0.8038&0.7286&0.7425/0.7436\\
&2.5D U-Net++&\textbf{0.8779}&\textbf{0.9627}&\textbf{0.9262}&\textbf{0.9310/0.9272}\\
\bottomrule
\bottomrule
\end{tabular}
}
\end{table}

\begin{figure}
\includegraphics[width=\textwidth]{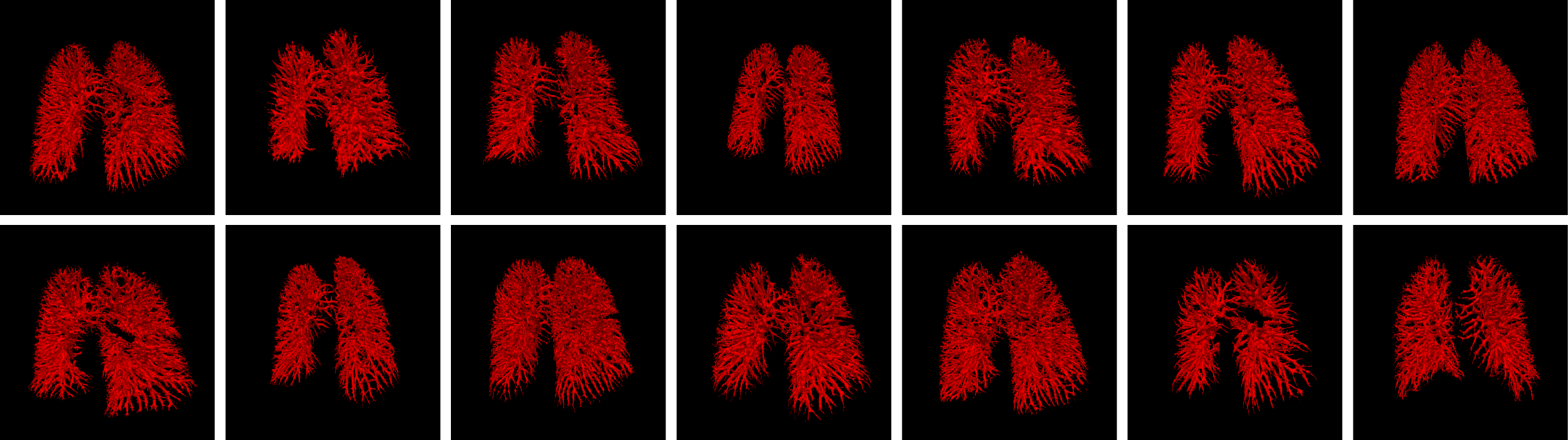}
\caption{Qualitative results of average fusion 2.5D U-Net++ on more CT images. Note that the position and shape of lung are varied from case to case, and the vessel segmentation results remain high performance on different quality CT images.} \label{fig3}
\end{figure}

Fig. \ref{fig2} shows the qualitative results of our methods compared with the ground truth. Results of more cases are displayed in Fig. \ref{fig3} to prove the robustness of model on different quality CT images. \\

\section{Conclusion}
Pulmonary Vessel Segmentation is one of the most challenging tasks in medical image analysis. The segmentation must overcome the complexity of the pulmonary structure as well as the limited resolution of CT images. In this paper, we propose a novel framework for automated pulmonary vessel segmentation based on a fused 2.5D convolution network structure. Slice radius is introduced to convolve adjacent information and the multi-planar fusion optimizes the presentation of intra and inter slice features. Besides, a post-processing is designed to refine the segmentation results using main components information of the pulmonary vessel tree. Our method excels others by a large margin on pulmonary vessel segmentation task and achieves very competitive results on DICE and Precision value.\\

\noindent\textbf{Acknowledgments.} This work is partially funded by Beijing Posdoctoral Research Foundation.
\bibliographystyle{splncs04}
\bibliography{miccai2019}

\begin{thebibliography}{10}
\providecommand{\url}[1]{\texttt{#1}}
\providecommand{\urlprefix}{URL }
\providecommand{\doi}[1]{https://doi.org/#1}

\bibitem{armato2011lung}
Armato, S.G., et~al.: The lung image database consortium (lidc) and image
  database resource initiative (idri): a completed reference database of lung
  nodules on ct scans. Medical physics  \textbf{38}(2),  915--931 (2011)

\bibitem{el2011lung}
El-Baz, A., Suri, J.S.: Lung imaging and computer aided diagnosis. CRC Press
  (2011)

\bibitem{kaftan2008fuzzy}
Kaftan, J.N., Kiraly, A.P., Bakai, A., Das, M., Novak, C.L., Aach, T.: Fuzzy
  pulmonary vessel segmentation in contrast enhanced ct data. In: Medical
  Imaging 2008: Image Processing. vol.~6914, p. 69141Q. International Society
  for Optics and Photonics (2008)

\bibitem{lesage2009review}
Lesage, D., Angelini, E.D., Bloch, I., Funka-Lea, G.: A review of 3d vessel
  lumen segmentation techniques: models, features and extraction schemes.
  Medical image analysis  \textbf{13}(6),  819--845 (2009)

\bibitem{li2018h}
Li, X., Chen, H., Qi, X., Dou, Q., Fu, C.W., Heng, P.A.: H-denseunet: Hybrid
  densely connected unet for liver and tumor segmentation from ct volumes. IEEE
  transactions on medical imaging  \textbf{37}(12),  2663--2674 (2018)

\bibitem{rudyanto2014comparing}
Rudyanto, R.D., Kerkstra, S., Van~Rikxoort, E.M., Fetita, C., Brillet, P.Y.,
  Lefevre, C., Xue, W., Zhu, X., Liang, J., {\"O}ks{\"u}z, {\.I}., et~al.:
  Comparing algorithms for automated vessel segmentation in computed tomography
  scans of the lung: the vessel12 study. Medical image analysis
  \textbf{18}(7),  1217--1232 (2014)

\bibitem{schneider2012tissue}
Schneider, M., Reichold, J., Weber, B., Sz{\'e}kely, G., Hirsch, S.: Tissue
  metabolism driven arterial tree generation. Medical image analysis
  \textbf{16}(7),  1397--1414 (2012)

\bibitem{shang2011vascular}
Shang, Y., et~al.: Vascular active contour for vessel tree segmentation. IEEE
  Transactions on Biomedical Engineering  \textbf{58}(4),  1023--1032 (2011)

\bibitem{shikata2004automated}
Shikata, H., Hoffman, E.A., Sonka, M.: Automated segmentation of pulmonary
  vascular tree from 3d ct images. In: Medical Imaging 2004: Physiology,
  Function, and Structure from Medical Images. vol.~5369, pp. 107--117.
  International Society for Optics and Photonics (2004)

\bibitem{simonyan2014very}
Simonyan, K., Zisserman, A.: Very deep convolutional networks for large-scale
  image recognition. arXiv preprint arXiv:1409.1556  (2014)

\bibitem{staal2004ridge}
Staal, J., Abr{\`a}moff, M.D., Niemeijer, M., Viergever, M.A., Van~Ginneken,
  B.: Ridge-based vessel segmentation in color images of the retina. IEEE
  transactions on medical imaging  \textbf{23}(4),  501--509 (2004)

\bibitem{tetteh2018deepvesselnet}
Tetteh, et~al.: Deepvesselnet: Vessel segmentation, centerline prediction, and
  bifurcation detection in 3-d angiographic volumes. arXiv preprint
  arXiv:1803.09340  (2018)

\bibitem{yun2019improvement}
Yun, J., Park, J., Yu, D., Yi, J., Lee, M., Park, H.J., Lee, J.G., Seo, J.B.,
  Kim, N.: Improvement of fully automated airway segmentation on volumetric
  computed tomographic images using a 2.5 dimensional convolutional neural net.
  Medical image analysis  \textbf{51},  13--20 (2019)

\bibitem{zhou2018unet++}
Zhou, Z., Siddiquee, M.M.R., Tajbakhsh, N., Liang, J.: Unet++: A nested u-net
  architecture for medical image segmentation. In: Deep Learning in Medical
  Image Analysis and Multimodal Learning for Clinical Decision Support, pp.
  3--11. Springer (2018)

\end{thebibliography}

\end{document}